\documentclass{cda}
\usepackage{array}

\begin{document}

\title{Sequências didáticas para o ensino de Astronomia utilizando o Stellarium}

\author{Adriano M. Oliveira$^1$}
\author{Cibele Kemeicik$^{2}$}
\author{Augusto C. T. Monteiro$^{1}$}
\author{Thalita S. Benincá$^{1}$}
\author{Carlos Daniel da S. Mattos$^{1}$}
\author{Guilherme L. Schmidt$^{1}$}

\instituto{$^1$Instituto Federal do Esp\'irito Santo (Ifes), Guarapari - ES, Brasil}
\instituto{$^2$Secretaria Municipal de Educação (Semed), Guarapari - ES, Brasil}

\abstract{The goal of this paper is to present three proposal of didactics sequence with astronomical thematic taking alignment with BNCC. They were developed during the work plan progress of students of high school connected with OAIG. As a final result of the students practices, they: (1) determined the mass of Jupiter by Io eclipse; (2) recognized crateras and seas of the Moon; and (3) constructed the H-R diagram for the near stars.}


\resumo{Esse artigo apresenta o resultado da aplicação de três sequências didáticas, alinhadas às habilidades e competências da BNCC, dentro da temática de Astronomia. Estas sequências foram desenvolvido durante a execução do plano de trabalho de IC-Jr (alunos de Ensino Médio) do Observatório Astronômico do Ifes Guarapari (OAIG) e resultaram na (o): (1) determinação da massa de Júpiter, calculada a partir do eclipse de Io; (2) reconhecimento das crateras e mares lunares; (3) construção do diagrama H-R para as estrelas mais próximas.}

\pchave{Ensino, Sequência Didática, Astronomia, Stellarium, Modelo Matemático.}

\maketitle

\oautor{Oliveira, A.M. et al}

\section{INTRODUÇÃO \label{sec:Int}}

A Base Nacional Comum Curricular (BNCC) \cite{bncc2018}, aprovada no ano de 2018, traz alguns desafios para os professores da educação básica. 
Dentre estes entraves, destaca-se o fato de que certos conteúdos trazidos pelo novo currículo não são trabalhados, ou são de forma superficial, durante a formação destes profissionais, como é o caso da Astronomia que, por sua vez, foi uma das áreas que mais ganhou espaço dentro desse novo currículo.

Uma análise das tabelas contidas nesse documento permite observar que houve uma distribuição de temas ligados ao estudo do céu em todo o currículo, seja de forma direta ou comunicando-se com outros conteúdos.
Desde o ensino fundamental, cujo tema é tratado dentro do eixo Ciências da Natureza, Terra e Universo, chegando até o ensino médio, cuja abordagem é feita dentro da relação entre Tecnologia e Ciências da Natureza, já que uma compreensão contemporânea do universo físico, da vida planetária e da vida humana está conectada ao entendimento dos instrumentos com os quais o ser humano maneja e investiga o mundo natural.
Nesse contexto, algumas habilidades e competências são desenvolvidas em momentos bem determinados. 
Por exemplo:
\begin{quote} \footnotesize
 {\it{(EF01CI05)} - } Identificar e nomear diferentes escalas de tempo: os períodos diários (manhã, tarde, noite), sucessão de dias, semanas, meses e anos;
\end{quote}
\begin{quote}\footnotesize
 {\it{(EF01CI06)} - } Selecionar exemplos de como a sucessão de dias e noites orienta o ritmo de atividades diárias de seres humanos e de outros seres vivos.
\end{quote}
A título de análise, o conjunto de letras e números dentro dos parênteses são os códigos das habilidades dentro da BNCC, o significado está ligado ao que deve ser abordado, em que momento isso acontece e a quem compete a abordagem, como é o caso da habilidade {\it{(EF01CI05)}}, que deve ser abordada no ensino fundamental {\it{(EF)}}, durante o primeiro ano {\it{(01)}}, pelo professor de ciências {\it{(CI)}} e essa é a quinta {\it{(05)}} habilidade a ser trabalhada. 
Outras citações desse tipo serão feitas dentro das sequências didáticas e têm a finalidade de destacar onde cada habilidade da BNCC é trabalhada. 
Assim, o ensino básico tem um currículo norteador, que garante aos estudantes o direito de aprender um conjunto de habilidades comuns e que está em concordância com o Plano Nacional de Educação (PNE) \cite{pne2015} e com a Lei de Diretrizes e Bases para a educação nacional (LDB) \cite{ldb1996}. 

Ainda, na linha dos desafios encontrados pelos professores, tem-se a aplicabilidade das habilidades, como é o caso das citadas acima, fato inerente às ciências da natureza. 
O alinhamento dessa aplicabilidade com o artigo 8$^o$ da BNCC aumenta o desafio deste profissional,
a saber:
\begin{quote} \footnotesize
Os currículos, coerentes com a proposta pedagógica da instituição ou rede de ensino, devem adequar as proposições da BNCC à sua realidade, considerando, para tanto, o contexto e as características dos estudantes, devendo IV. Conceber e pôr em prática situações e procedimentos para motivar e engajar os estudantes nas aprendizagens \cite[art. 8]{bncc2018}.
\end{quote}

Nesse ponto, em particular, a implementação de metodologias ativas pode ser uma forma de adequar a realidade escolar aos documentos que regem a educação básica. Apesar deste artigo não utilizar tal metodologia, as sequências didáticas, aqui apresentadas, podem ser adaptadas a ela, seguindo o proposto em \cite{carvalho2018}.

Com isso em mente, durante a execução dos planos de trabalho (PT) dos alunos de IC-Jr do Observatório Astronômico do Ifes Guarapari (OAIG), três tópicos foram selecionados e transformados em sequência didática (SD), a saber: determinação da massa de Júpiter, reconhecimento das crateras e mares lunares e a construção de um diagrama H-R. 
Aqui, apresentaremos uma sistematização das atividades utilizando o Stellarium como ferramenta para a coleta de dados.

\section{A OBSERVAÇÃO DO CÉU NOTURNO:  JÚPITER, LUA E CONSTELAÇÕES}

Revolucionando não só o modo como a humanidade enxergava o universo, mas também a forma de compreender a ciência no século XVII, Galileu Galilei (1564-1642) foi um astrônomo renascentista que, dispondo de uma luneta para a observação celeste, obteve evidências para a comprovação do modelo heliocêntrico proposto por Nicolau Copérnico. 
Por meio do aperfeiçoamento do telescópio, criação atribuída ao oculista Hans Lippershey (1570-1619), Galileu pôde empregá-lo em suas observações atreladas a extraordinárias descobertas que viriam mudar a concepção a qual o ser humano detinha sobre o cosmos.
Utilizando do princípio descoberto por Lippershey, que a combinação de lentes a uma determinada distância promovia o aumento de objetos distantes, o astrônomo italiano conseguiu aprimorar a capacidade de aproximação em três vezes, e, posteriormente, construiu instrumentos com um potencial trinta vezes maior\footnote{Um texto mais detalhado pode ser encontrado em \url{http://www.if.ufrgs.br/mpef/mef008/aulas_11/Galileu_observacoes_tel_v3.htm}.},  citada pelo filósofo Alexandre Koiré:
\begin{quote}
Montanhas na lua, novos “planetas” no céu, novas estrelas fixas em número tremendo, coisas que nenhum olho humano havia jamais visto e que nenhuma mente humana havia concebido anteriormente. E não só isso: além desses fatos novos, estarrecedores e inteiramente inesperados e imprevistos, havia ainda a descrição de uma invenção assombrosa, a do perspicillium\footnote{Perspicillium foi a denominação dada por Galileu ao seu aparato astronômico, conhecido atualmente como luneta.}, um instrumento - o primeiro instrumento científico - que havia tornado estas descobertas possíveis e possibilitado a Galileu transcender a limitação imposta pela natureza - ou por Deus - aos 
sentidos e ao conhecimento humano \cite[p.81]{koire2006mundo}.
\end{quote}

Em 1610, as descobertas astronômicas de Galileu Galilei são publicadas na obra denominada "Sidereus Nuncius" (O Mensageiro Sideral), nela são expostos as irregularidades da superfície lunar e alguns aglomerados de estrelas, além do aspecto mais importante da obra, a evidenciação de quatro luas orbitando Júpiter. A partir do seu trabalho, abriu-se a possibilidade de refutação do modelo ptolomaico, uma vez que, ali, desenvolveram-se argumentos que reafirmaram as conclusões que Nicolau Copérnico havia defendido em {\it{``De Revolutionibus Orbium Caelestium''}} (Sobre a revolução das esferas celestes), ou seja, a Terra era parte de um sistema planetário em que eram descritas órbitas ao redor do Sol.

\begin{figure}[!hbtp]

\begin{center}

\includegraphics[scale=0.5]{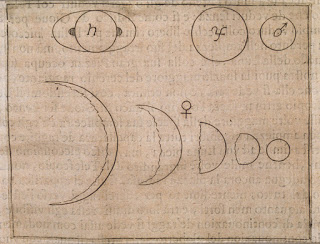}
\end{center}
\caption{Representação das observações feitas por Galileu, publicadas na obra  ``O Experimentador''.  Na parte superior, da esquerda para a direita, temos: (a) Saturno e seus anéis,  ilustrado como três esferas; (b) Júpiter e (c) Marte. Na parte inferior estão ilustradas as fases de Vênus. Fonte: \url{ www.astronomy2009.org}}

\label{Fig:OrelhaSaturno}
\end{figure}

A observação das fases de Vênus foi uma das evidências que colaboraram com o declínio do modelo ptolomaico.
Galileu, usando sua invenção, notou que Vênus têm fases, semelhante ao que acontece na Lua. 
As conclusões obtidas a partir disso, diz respeito aos planetas não apresentarem luminosidade própria, já que brilham quando recebem luz solar e que as mudanças aparentes das porções visíveis de Vênus somente poderiam ser explicadas caso fosse admitido que o planeta realizasse uma trajetória ao redor do Sol. 
Ademais, empregando sua luneta também conseguiu observar os anéis de Saturno, todavia, como o equipamento não tinha a resolução necessária para que a estrutura fosse vista nitidamente, Galileu descreveu o planeta de modo que fosse formado por três esferas, com uma lua bem próxima em cada lado, intitulando de “planeta com uma par de orelhas", como mostra a figura (\ref{Fig:OrelhaSaturno}).

\subsection{Determinando a Massa de Júpiter}
\begin{figure}[!hbtp]

\begin{center}

\includegraphics[scale=0.5]{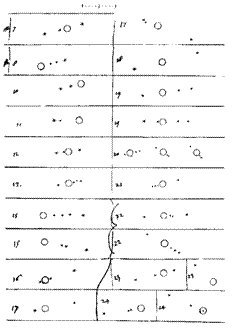}
\end{center}
\caption{Representação da posição das luas de Júpiter. Fonte: \url{http://www.if.ufrgs.br/mpef/mef008/aulas_11/Galileu_observacoes_tel_v3.htm}}
\label{Fig:RepLuasJupiter}
\end{figure}

Indubitavelmente, o aspecto mais significativo da obra de Galileu foi a importante descoberta de quatro estrelas errantes movendo-se em torno de Júpiter. 
A partir disto, foi realizado um estudo e acompanhamento detalhado das posições dessas estruturas ao redor do gigante gasoso, representando-os através de figuras (\ref{Fig:RepLuasJupiter}) e atribuindo a designação de planetas mediceanos, em homenagem ao grão-duque da Toscana, de modo que a posteriori passaram a ser consideradas satélites naturais e nomeadas de: Io, Ganimedes, Calisto e Europa. 
À vista disso, a evidenciação de luas orbitando outro planeta mostrou que o movimento dos corpos celestes não aconteciam somente em torno da Terra, como previa o modelo geocêntrico, assim, mesmo que indiretamente, são obtidos indícios que comprovam o sistema copernicano.

Uma discussão acerca dessas observações com simulações, usando as datas e horários históricos, bem como a relação do movimento das luas observado por Galileu e um paralelo com o movimento harmônico simples, pode ser encontrada em \cite{cuzinatto2014observaccoes}. 
Seguindo essa linha, utilizaremos a determinação da massa do planeta Júpiter como um dos temas de pesquisa da SD. 
Dentro desse contexto,  traçaremos uma estratégia para conduzir o aluno durante sua pesquisa, dividindo a temática em etapas.

\begin{itemize}
    \item {\bf{J1 - Caracterização das estruturas do sistema solar}}
    
    Iniciamos essa abordagem partindo do conhecimento dos alunos acerca das estruturas que compõe o Sistema Solar. 
    Perguntas sobre os planetas, luas, asteroides e outros corpos devem ser realizadas.
    Além disso,  podem ser incentivadas pesquisas sobre o tamanho, a massa, o movimento e outras características dessas estruturas.
    Para auxiliar nessa abordagem, a referência \cite{madejsky2014curso} pode ajudar.
    Em sala de aula, as informações coletadas pelos alunos podem ser utilizadas para mostrar que o Sol detém 99,86 $\%$ da massa total do Sistema Solar e que  Júpiter tem uma massa ($M_J$) maior que o dobro das soma das massas ($M$) dos demais planetas, $M_J = 2,5 M$.
    
    Cumprida essa etapa, o aluno conseguirá identificar os planetas e os corpos menores - tais como: planetas anões, luas, meteoros e cometas. 
    Para conduzir esse estudo teórico, pode ser utilizado o livro do \cite{oliveira2004}\footnote{Uma versão virtual da referência \cite{oliveira2004} está disponível no endereço \url{http://astro.if.ufrgs.br/}}.
    
    Durante a execução dessa tarefa o professor trabalhará as seguintes habilidades da BNCC:
    
     {\it{EF09CI14} -} Descrever a composição e a estrutura do Sistema Solar (Sol, planetas rochosos, planetas gigantes gasosos e corpos menores), assim como a localização do Sistema Solar na nossa Galáxia (a Via Láctea) e dela no Universo (apenas uma galáxia dentre bilhões);

    {\it{EF09CI16} -} Selecionar argumentos sobre a viabilidade da sobrevivência humana fora da Terra, com base nas condições necessárias à vida, nas características dos planetas e nas distâncias e nos tempos envolvidos em viagens interplanetárias e interestelares;
    
    {\it{EF09CI15} -} Relacionar diferentes leituras do céu e explicações sobre a origem da Terra, do Sol ou do Sistema Solar às necessidades de distintas culturas (agricultura, caça, mito, orientação espacial e temporal etc). 
    
    \item {\bf{J2 - Observação dessas estruturas}}
    
    Nessa etapa, iniciamos a localização das estruturas celestes por meio da observação do céu noturno. 
    Em sala de aula, a utilização de simuladores de céu, como o Stellarium\footnote{O Stellarium é um programa que simula o céu e sua versão para computador pode ser baixado na página \url{www.stellarium.org}, gratuitamente.}, torna-se uma ferramenta essencial. 
    Além disso, o uso de cartas celestes\footnote{A consulta ou impressão das cartas celestes podem ser feitas através dos sites \url{www.cartascelestes.com} ou \url{https://observatorio.guarapari.ifes.edu.br/index.php/planisferio/}.} pode ampliar a capacidade de abstração, localização, posicionamento e leitura.
    Esse trabalho de localização dos planetas nos levará a conclusões naturais como a divisão do céu em constelações, extrapolando o objetivo inicial da prática e alinhando-se com a BNCC na habilidade
    
    {\it{EF03CI08} -} Observar, identificar e registrar os períodos diários em que o Sol, demais estrelas, Lua e planetas estão visíveis no céu.
    
    \item {\textbf{J3 - Contribuições de Galileu para a Astronomia}}
    
   Aqui, buscamos a compreensão do processo de evolução do conhecimento científico, no tocante às observações de Galileu e sua contribuição para a Astronomia. 
   Nesse sentido, pode-se incentivar a construção de lunetas semelhantes à de Galileu. 
   Alguns modelos podem ser encontrados nos trabalhos \cite{cardosoluneta,daconstruccao,oliveira2010luneta}\footnote{Uma sugestão de vídeo ensinando a construir a luneta pode ser visto em \url{www.youtube.com/watch?v=quP7pOORCv0}.}. 
   Essa etapa vai ao encontro da BNCC na habilidade destacada abaixo:
   
   {\it{EF05CI13} -} Projetar e construir dispositivos para observação à distância (luneta, periscópio etc.), para observação ampliada de objetos (lupas, microscópios) ou para registro de imagens (máquinas fotográficas) e discutir usos sociais desses dispositivos.

  O uso das lunetas, confeccionadas pelos alunos, deve ser incentivado na prática de observação noturna. 
  Em sala de aula, o uso do Stellarium para mostrar as estruturas observadas tanto por Galileu quanto pelos estudantes ajudará no reconhecimento do céu.
  Nesse momento, vale mostrar que a trajetória dos planetas, como a observamos, está quase sobre o caminho percorrido pelo Sol, a este damos o nome de Eclíptica.
  Outras curiosidades e problemas dos modelos com órbita circular podem ser levantados, por exemplo, a velocidade orbital variável e o movimento retrógrado dos planetas.
  De fato, o movimento retr\'ogrado \'e uma consequ\^encia do movimento relativo dos planetas, que apresentam velocidades diferentes em suas \'orbitas em torno do Sol.
  Ele pode ser explicado considerando epiciclos nas \'orbitas circulares, como feito por Cop\'ernico, ou utilizando o modelo de \'orbitas el\'ipticas, como proposto por Kepler. Esse último simplifica o modelo matemático e descreve as órbitas planetárias de forma precisa.
  
   \item {\textbf{J4 - Dinâmica Planetária}}  
  
    Solicitamos aos alunos que modelassem o movimento planetário observado no Stellarium e verificamos o quão próximo dos modelos aceitos pela ciência eles se aproximaram. 
      Além disso, pode ser um bom momento para trabalhar os modelos Geocêntricos e Heliocêntricos, chegando até as Leis de Kepler e a Lei da Gravitação Universal de Newton, o livro \cite{rooney2013historia} pode dar elementos para a discussão.
   
    Devido à relevância das Leis de Kepler para esse trabalho, e buscando ampliar a abrangência do que está aqui exposto, optamos por enunciar tais leis, cuja síntese pode ser encontrada em \cite{garms2018}, dando o devido destaque a alguns pontos onde os alunos comumente cometem erros.
   
 {\textbf{1$^a$ Lei de Kepler - Lei das Órbitas}}
   
    A órbita planetária é um elipse, com o Sol ocupando um dos focos. 
    Contudo, o valor da excentricidade dela é muito próximo a zero, o que a aproxima  a uma circunferência.
    Tal característica é mostrada na tabela (\ref{Tab:Excentricidade}), a qual traz as excentricidades das órbitas planetárias, e ilustrada na figura (\ref{Fig:Excentricidade}). 
    Essas figuras foram obtidas utilizando os valores da excentricidade ($e$) de cada planeta e a relação desta com os semieixos maior ($a$) e menor ($b$), dada por
    \begin{eqnarray}
         a = c/e\quad\text{e}\nonumber\\
         b = a\sqrt{1-e^2}\quad,
    \end{eqnarray}
    sendo $2c$ a distância entre os focos da elipse. 
    Dispondo dos valores dados pela relação acima e usando as equações paramétricas da elipse,
    \begin{eqnarray}
         x = a \cos(t) \quad \text{e}\nonumber\\
         y = b \sin(t)\quad,
    \end{eqnarray}
    teremos a forma das órbitas planetárias como mostrado nas figuras (\ref{Fig:Excentricidade}).
    Vale destacar que as órbitas estão fora de escala e os eixos foram relativizados.
    Isso facilita a comparação das órbitas e mostra o quão próximas de uma circunferência elas são.
    
\begin{table}[!hbtp]
\centering
\caption{Essa tabela mostra o valor da excentricidade da órbita de cada um dos planetas do Sistema Solar, \cite{madejsky2014curso}.}
\label{Tab:Excentricidade}
\begin{tabular}{c|c}
\hline
{\textbf{Planeta}}  & {\textbf{Excentricidade}} \\
\hline
Mercúrio & 0,206 \\ \hline
Vênus & 0,007\\ \hline
Terra & 0,017\\ \hline
Marte & 0,093\\ \hline
Júpiter & 0,049\\ \hline
Saturno & 0,056\\ \hline
Urano & 0,046 \\ \hline
Netuno & 0,011\\ \hline
\end{tabular}
\end{table}

\begin{figure}[!hbtp]
\begin{center}

\includegraphics[scale=0.4]{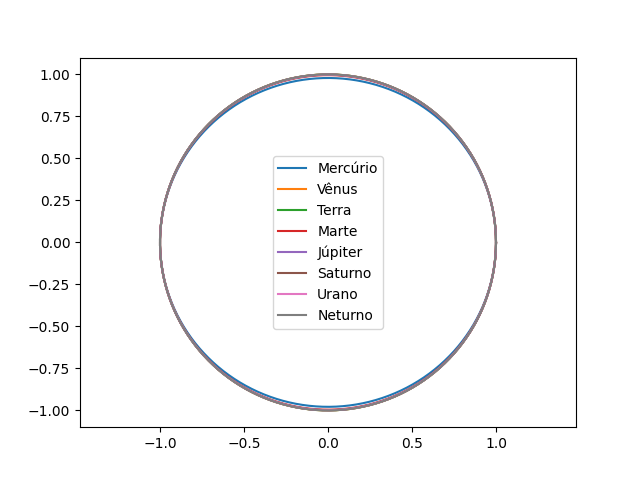}
\includegraphics[scale=0.4]{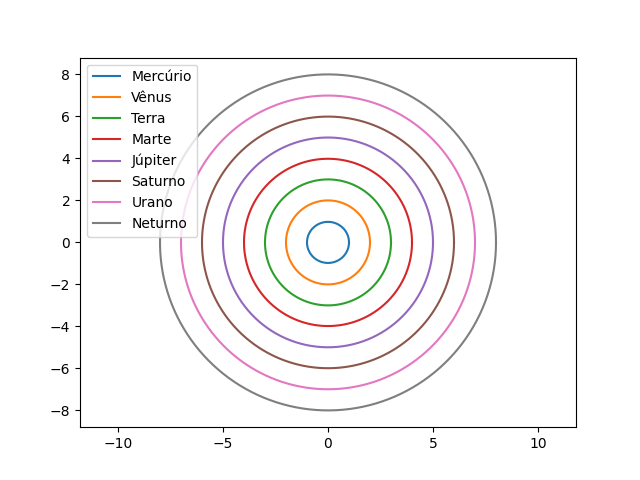}

\end{center}
\caption{Nessa figura é mostrada uma comparação entre as excentricidades das órbitas planetárias em duas perspectivas. Na superior, as órbitas são apresentadas com o mesmo valor de $a$, que adotamos igual a unidade. Na figura inferior, tais órbitas foram distanciadas de uma constante. Todas as órbitas são muito próximas a uma circunferência devido ao pequeno valor da excentricidade das elipses representadas.}
\label{Fig:Excentricidade}
\end{figure}

\textbf{2$^a$ Lei de Kepler - Lei das Áreas}
  
    Ao imaginarmos uma linha ligando algum planeta ao Sol, essa linha varrerá áreas iguais em intervalos de tempos iguais, independente da posição desse planeta.
    Em outras palavras, a velocidade areolar ($v_{\text{areolar}}$) -- rapidez com que uma área é varrida, pela linha imaginária -- é constante. 
    Então,
    \begin{equation}
        v_{\text{areolar}}=\frac{\Delta A}{\Delta t} = \text{constante}\quad,
    \end{equation}
    sendo $\Delta A$ a variação da área e $\Delta t$ o intervalo de tempo em que a área é varrida.
    Portanto, da figura (\ref{Fig:2LeiKepler}), se o tempo para o planeta ir de A até B ($\Delta t_{AB}$) é igual ao tempo que o planeta leva para ir de C até D ($\Delta t_{CD}$), então $\Delta A_{AB} = \Delta A_{CD}$.
    \begin{figure}[!hbtp]
    \begin{center}

    \includegraphics[scale=0.3]{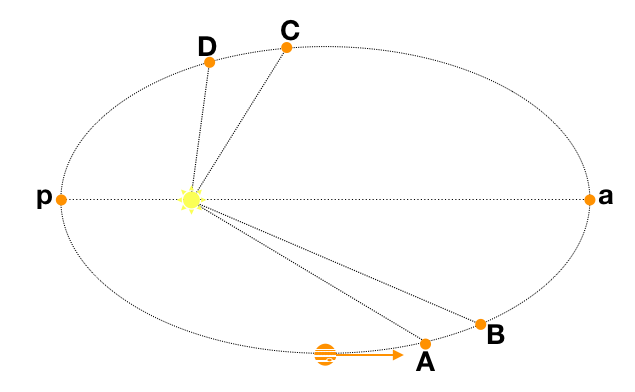}

    \end{center}
    \caption{Essa figura é uma representação fora de escala (a excentricidade está muito grande) de uma órbita planetária. Nela estão marcados os pontos de afélio (a) e periélio (p), além de quatro pontos aleatoriamente escolhidos A, B, C e D.}
    \label{Fig:2LeiKepler}
    \end{figure}
    
    Contudo, como a órbita mostrada na figura é uma elipse, o tamanho da linha imaginária que liga o Sol ao planeta muda com o tempo. 
    Isso faz com que o espaço linear percorrido pelo planeta não seja o mesmo em intervalos de tempos iguais. 
    Assim, a distância entre os pontos A e B é menor que a distância entre os pontos C e D, $\Delta S_{AB} < \Delta S_{CD}$, uma vez que a linha imaginária cresce quando o planeta se afasta do Sol.
    Desse modo, a velocidade tangencial próxima ao afélio (ponto 'a', da figura (\ref{Fig:2LeiKepler})) é menor que a velocidade próxima ao periélio (ponto 'p', nessa mesma figura).
    Consequentemente, quando o planeta se aproxima do periélio sua velocidade tangencial deve aumentar, logo ele experimenta, na direção tangencial, um movimento acelerado, já quando se aproximar do afélio experimentará um movimento retardado, na mesma direção. 
    
    \textbf{3$^a$ Lei de Kepler - Lei do Período}

    Por fim, a razão entre o quadrado do período de translação ($T$) de algum planeta e o cubo do semieixo maior ($a$) deste é constante ($K$), ou seja,
    \begin{equation}
    \frac{T^2}{a^3} = K\quad.
    \label{eq:3Lei}
    \end{equation}
    Entretanto, a excentricidade da órbita planetária é próxima a zero, o que faz $a$ se aproximar ao raio de uma circunferência. 
  
  \item {\textbf{J5 - Coleta de dados}}

Uma vez que as leis de Kepler são válidas para qualquer sistema orbitante, pode-se determinar a massa de Júpiter utilizando (\ref{eq:3Lei}). 
Como a excentricidade orbital de Io é pequena, algo em torno de 0,0041, podemos aproximar o movimento de revolução desta lua ao redor de Júpiter ao movimento circular uniforme.
Isso facilitará a abordagem matemática sem mudar significativamente a massa de Júpiter, determinada pelo estudante.

Sendo assim, como a formulação da Lei da Gravitação Universal de Isaac Newton, que unifica as Leis de Kepler,  expressa a interação entre dois corpos isolados, de modo que a força de atração gravitacional ($F_g$) entre eles é diretamente proporcional ao produto das massas ($M$ e $m$) e inversamente proporcional ao quadrado da distância ($r^2$),  então
\begin{equation}
    F_g = G\frac{M m}{r^2}\quad,
    \label{eq:gravuni}
\end{equation}
sendo $G=6,674184\times 10^{-11}\quad m^{3}kg^{-1}s^{-2}$ uma constante de proporcionalidade, chamada de constante universal da gravitação.
No sistema Júpiter-Io, $M$ torna-se a massa do planeta, $m$ a massa de Io e $r$ o raio orbital da lua.
Como a única força que atua sobre a lua é a gravitacional, então a força resultante
\begin{equation}
    F_R = m a
\end{equation}
é centrípeta, para um referencial inercial centrado em Júpiter.
Desse modo, substituindo a aceleração centrípeta ($a = \omega^2 r$) na equação acima, em que $\omega = 2\pi/T$ é a velocidade angular de Io e $T$ é o período orbital da lua, e desenvolvendo-a temos
\begin{equation}
    \frac{T^2}{r^3} = \frac{4\pi^2}{GM}\quad,
    \label{eq:NewKep}
\end{equation}
como enunciado por Kepler em sua terceira lei.
Por outro lado, a demonstração do resultado acima quando $M \approx m$ será deixado como  desafio, o resultado para esse caso pode ser encontrado em \cite{oliveira2004}.
De todo modo, se conhecermos o período orbital e o raio da órbita de Io, conseguiremos determinar a massa aproximada de Júpiter. 
Assim, para coletar esses dados utilizamos o aplicativo Stellarium seguindo pois o passo-a-passo abaixo:

(i) Retiramos os efeitos de atmosfera e de superfície;

(ii) Fixamos o programa para observar apenas Júpiter, visto da Terra.
Isso pode ser feito localizando o planeta e clicando sobre ele;

(iii) Aceleramos o tempo até que Io estivesse em uma situação de saída do eclipse e anotamos o instante (data e horário) em que isso ocorre. Tal estudo pode ser realizado no período de um mês, anotando as datas e horários em que Io se encontra nessa situação.

(iv) Configuramos a localização de observação para Júpiter;

(v) Fixamos o programa para observar apenas Io, visto na superfície de Júpiter.
Isso pode ser feito localizando a lua e clicando sobre ela;

(vi) Usando a função de data e hora, selecionamos os dias e horários anotados no item (iii) e tomamos nota da distância que Io está de um observador, localizado na superfície de Júpiter, em cada uma delas. 
Vale ressaltar que repetimos esse procedimento até obtermos as distâncias correspondentes a cada instante anotado.

Ao fim da coleta de dados, obtivemos a tabela (\ref{Tab:Io}). 
Ela nos mostra os dias e horários em que Io está na situação descrita no item (iii) e a distância coletada no item (vi).
\begin{table}[!hbtp]
\centering
\caption{Essa tabela mostra as horas que Io sai do eclipse, visto da Terra, e as distâncias que a referida lua está de Júpiter.}
\label{Tab:Io}
\begin{tabular}{c|c|c}
\hline
{\textbf{Data}}  & {\textbf{Hora}}&{\textbf{Distância (km)}} \\
\hline
16/06/2020 & 21:47:14 & 455640,398 \\ \hline
18/06/2020 & 16:12:17 & 470879,512 \\ \hline
20/06/2020 & 10:24:17 & 379215,874 \\ \hline
22/06/2020 & 04:53:25 & 382994,694 \\ \hline
23/06/2020 & 23:15:34 & 475846,748 \\ \hline
25/06/2020 & 17:37:47 & 448806,513 \\ \hline
27/06/2020 & 11:58:15 & 358118,779 \\ \hline
29/06/2020 & 06:28:45 & 417394,157 \\ \hline
01/07/2020 & 00:57:54 & 483537,748 \\ \hline
02/07/2020 & 20:08:07 & 387340,763 \\ \hline
04/07/2020 & 13:55:20 & 362064,246 \\ \hline
06/07/2020 & 08:22:40 & 458900,405 \\ \hline
08/07/2020 & 03:03:03 & 463288,116 \\ \hline
09/07/2020 & 21:35:39 & 363872,101 \\ \hline
11/07/2020 & 15:53:50 & 399555,370 \\ \hline
13/07/2020 & 09:53:39 & 438333,768 \\ \hline
15/07/2020 & 04:39:48 & 432320,230 \\ \hline
\end{tabular}
\end{table}

Essa tabela apresenta uma diferença significativa nas distâncias medidas.
Tal diferença está ligada ao tamanho do planeta e a rapidez rotacional dele, já que ao escolhermos um referencial em Júpiter, ele é posto sobre a superfície do planeta, na mesma latitude em que o observador se encontrava na superfície da Terra, e não em seu centro\footnote{Note que foi mudada apenas a localização do observador para o planeta Júpiter. Desse modo, no Stellarium, não alteramos as coordenadas de latitude e longitude, o que coloca o observador em Júpiter na mesma latitude em que estava na Terra. Ou seja, se, originalmente, o observador estava em Guarapari-ES, ele ficará localizado na latitude de cerca de 20 graus ao sul do equador da Terra e também em Júpiter.}. 
Durante a coleta de dados utilizamos dois referenciais: o referencial A, está na Terra e observa o eclipse de Io, definido no passo (ii) anterior, este referencial mede os intervalos de tempo, e o referencial B, está sobre a superfície de Júpiter medindo as distâncias entre ele e Io, nos instantes estabelecidos por A. 
Apesar de podermos aproximar a posição de A para o centro da Terra, já que a distância entre a Terra e Júpiter é muito maior que o raio da Terra. 
O mesmo não pode ser feito para B, ou seja, a distância entre B e Io será influenciada pela rotação de Júpiter, de modo que  B se aproxima e se afasta de Io, especialmente porque o observador está localizado numa latitude baixa, próxima ao equador do planeta, causando as mudanças nas distâncias apresentadas na tabela (\ref{Tab:Io}). Note que, se o observador estivesse em um dos polos de Júpiter, não haveria esse efeito de variação da distância produzido pela rotação do planeta.

Quando calculamos a diferença entre os valores máximo e mínimo apresentados, $\approx 1,3\times 10^{5}$ km, e comparamos com o diâmetro de Júpiter, $\approx 1,43\times 10^{5}$ km, vemos que os valores são muito próximos, evidenciando o que foi exposto acima.
Sendo assim, para se obter uma estimativa, aproximada e razoável, da distância entre Io e o centro de Júpiter, uma vez que a latitude influencia no valor obtido para o raio orbital da lua, é necessário calcular a média a partir de uma ``grande'' quantidade de dados coletados (no nosso caso, durante um mês).

Além disso, durante a coleta de dados foram desconsiderados os movimentos de translação dos planetas (movimento em torno do Sol), ou seja, tratamos o caso onde os dois planetas, Terra e Júpiter, estão em repouso (são referenciais inerciais).
Apenas com essa hipótese, os tempos medidos por A correspondem exatamente ao período sideral de Io.

\item {\textbf{J6 - Obtendo a Massa de Júpiter}}

A partir dos dados expostos na tabela acima, encontramos o período orbital médio de Io, o qual é  
$$<T> = (42,5 \pm 0,3) \; \text{horas.}$$
Para isso, calculamos a diferença entre dois intervalos de tempo sucessivos.
Esse valor obtido depende da velocidade tangencial relativa entre a Terra e Júpiter. 
Quando Júpiter está em oposição, em 2020 esse fenômeno ocorreu no dia 14 de julho, próximo a data de coleta dos dados, essa diferença de tempo corresponderá, com maior precisão, ao período orbital correto.

Na situação de oposição a velocidade tangencial da Terra é praticamente paralela a velocidade orbital de Júpiter, ou seja, a posição relativa entre os planetas não muda muito entre dois eclipses sucessivos de Io, o que permite obter uma melhor estimativa do período orbital real da lua.
Entretanto, se nossos dados fossem coletados quando Júpiter estivesse em uma situação de quadratura, ou seja, para o caso onde as velocidades tangenciais dos planetas em suas órbitas são perpendiculares, o que acontece nas épocas próximas das quadraturas de Júpiter, teríamos um aumento ou diminuição na medida do período orbital de Io, devido à velocidade finita de propagação da Luz.
Isso está diretamente relacionado com a primeira determinação quantitativa da velocidade da luz, realizada no século XVII, por Ole Romer\footnote{Para maiores detalhes sobre esse fato histórico, consulte \url{https://en.wikipedia.org/wiki/Ole_R\%C3\%B8mer}.}.

Do mesmo modo, encontramos que o raio médio da órbita de Io é   
$$<r> = (42 \pm 4) \times 10^4 \; \text{km.}$$

Para calcular o erro dos resultados apresentados acima, utilizamos o cálculo de desvio padrão médio ($<\sigma>$),
\begin{equation}
<\sigma> = \sqrt{\frac{\sum_{i=1}^{N}(x_i - <x>)^2}{N-1} }\quad,  
\label{eq:DesvPadrao}
\end{equation}
tal que $N$ é o número de elementos, $x_i$ é o valor da $i$-ésima grandeza e $<x>$, o valor médio desta.

Portanto, substituindo os valores de $<T>$ e $<r>$ na equação (\ref{eq:NewKep}), obteremos a massa de Júpiter 
$$
M_J = (1,9 \pm 0,5)\times 10^{27}\quad kg\quad,
$$
cujo desvio foi calculado usando a expressão de propagação do erro
\begin{eqnarray}
dX(x_i) = \sum \left(\frac{\partial X}{\partial x_i}\right) dx_i\quad,
\label{eq:PropErro}
\end{eqnarray}
que ao ser aplicada na equação (\ref{eq:NewKep}) e desconsiderando os termos de ordem superior, resulta em
$$
\bigg(\frac{\Delta M_J}{M_J}\bigg)^2\approx\bigg(3\frac{\Delta r}{r}\bigg)^2 + \bigg(2\frac{\Delta T}{T}\bigg)^2 \quad,
$$
sendo $\Delta M_J$, $\Delta r$ e $\Delta T$ os erros das grandezas correspondente a $M_J, r$ e $T$, respectivamente.
O valor encontrado para a massa de Júpiter está de acordo com o resultado esperado e apresentado na literatura, algo em torno de $1,898 \times 10^{27} kg$.

Vale destacar que esse método pode ser utilizado para encontrar a massa de um corpo central, quando for possível obter estimativas do período e do semieixo maior da órbita do corpo menor em torno dele. Uma aplicação bem interessante desse método é o da determinação da massa do buraco negro supermassivo que há no centro da nossa galáxia a partir da observação da órbita de estrelas que estão próximas a ele, como vem sendo feito recentemente.
\end{itemize}

\subsection{As Crateras e Mares Lunares}

Quando Galileu apontou o telescópio para a Lua, ele observou que sua superfície apresenta um relevo com montanhas e vales similares aos encontrados na Terra, contrariando a concepção aristotélica do satélite ser perfeitamente esférico.
Diante dessas provas observacionais, a noção de que o meio celeste seria perfeito, imutável e eterno não encontra argumentos que possam justificá-lo, e a divisão em mundo sublunar e supralunar é abandonada, já que, contrariando o que se acreditava, as estruturas siderais apresentam imperfeições e estão suscetíveis a mudanças.
De fato, a incoerência desse conceito fica ainda mais evidente quando o astrônomo observa as manchas solares, que não possuem uma forma fixa e existência permanente.

Nesse sentido, essa SD tem o intuito de reconhecer as crateras e mares lunares, como instrumento para adquirir algumas das habilidades e competências previstas pela BNCC.
Desse modo, seguiremos a linha de raciocínio proposta na seção anterior e dividiremos o percurso nas seguintes metas:
\begin{itemize}
    
    \item \textbf{L1 - Observação da Lua}
   
   Para a abordagem inicial dessa sequência didática, buscamos incentivar a observação da Lua, feita sem instrumentos astronômicos, com o objetivo de aumentar a interação dos estudantes com a natureza que nos cerca.
    A partir dessa observação sistemática do astro, algumas conclusões  poderão ser tiradas, por exemplo: o atraso de, aproximadamente, 50 (cinquenta) minutos em seu horário de nascimento, quando comparamos dois dias consecutivos, o efeito dela sobre os fluidos da Terra e as manchas escuras em sua superfície.
    No entanto, com maior tempo observacional, outros detalhes acerca da trajetória lunar poderão ser percebidas, como é o caso da diferença entre os caminhos percorridos pelo Sol e pela Lua, que não são exatamente os mesmos. 
    Cabe frisarmos que, neste caso mais detalhado, poderá haver ainda uma discussão acerca da condição particular para a ocorrência de eclipses, os quais ocorrem apenas quando a Lua está cheia ou nova e o Sol está sobre a linha dos nodos\footnote{A linha dos nodos é uma linha formada pela intersecção do plano da órbita da Terra em torno do Sol com o plano da órbita da Lua em torno da Terra \cite{oliveira2004}.}.
    
    Em sala de aula, todos esses aspectos foram discutidos e mostrados utilizando o Stellarium.
    Nesta abordagem com o {\it{software}}, escolhendo quatro dias, um para cada fase da Lua, evidenciamos  aos alunos a variação da porção iluminada do satélite,  uma vez que neste ocorre mudança de área iluminada ao passar dos dias. 
    Da mesma forma, aproveitamos esse momento para compararmos, no planetário virtual, as posições do Sol e da Lua nos dias em que os dois astros estão visíveis no céu simultaneamente.
    Portanto, percebemos a diferença de trajetória entre eles e, a partir dai, fizemos um paralelo deste caso com datas em que o eclipse ocorre. 
    Assim, abordamos duas habilidades previstas pela BNCC:

    {\it{EF05CI12} -} Compreender a periodicidade das fases da Lua;

    {\it{EF08CI12} -} Justificar, por meio da construção de modelos e da observação da Lua no céu, a ocorrência das fases da Lua e dos eclipses, com base nas posições relativas entre Sol, Terra e Lua.
    
    \item \textbf{L2 - Astrofotografia e a Identificação das Crateras e dos Mares Lunares}
    
    Em uma das nossas observações, realizamos capturas de imagens da Lua utilizando equipamentos do Observatório Astronômico do Ifes Guarapari (OAIG). 
    Por intermédio de um smartphone acoplado no telescópio Meade LX-90 (12 polegadas), obtivemos as figuras (\ref{Fig:Lua1}, \ref{Fig:Lua2} e \ref{Fig:Lua3}) que nos possibilitaram, através de comparação com documentos disponíveis\footnote{Os padrões utilizados para comparação podem ser encontrados em: \url{www.bbc.co.uk/programmes/articles/5h42ZxMX9QW7hRvlVC7RKbn/section-2-bright-and-dark-craters} e \url{ www.bbc.co.uk/programmes/articles/5gdrKwHtXhRkcq0xHDdhqvj/section-1-the-lunar-seas.}}, reconhecer e caracterizar $10$ (dez) mares e $27$ (vinte e sete) crateras da face visível do satélite. 
    Não obstante, é necessário ressaltar que o uso do Stellarium é, também, uma ótima ferramenta alternativa para identificar essas características.
    Este reconhecimento no {\it{software}} pode ser feito fixando a observação na Lua e ampliando a imagem para, assim, tais irregularidades ficarem mais evidentes. 
    Nesse sentido, é  interessante coletar essas imagens em fases diferentes da Lua para que se consiga observar e identificar um maior número de detalhes, devido ao maior contraste na imagem.
    
    \begin{figure}[!hbtp]
        \begin{center}
            \includegraphics[scale=0.23]{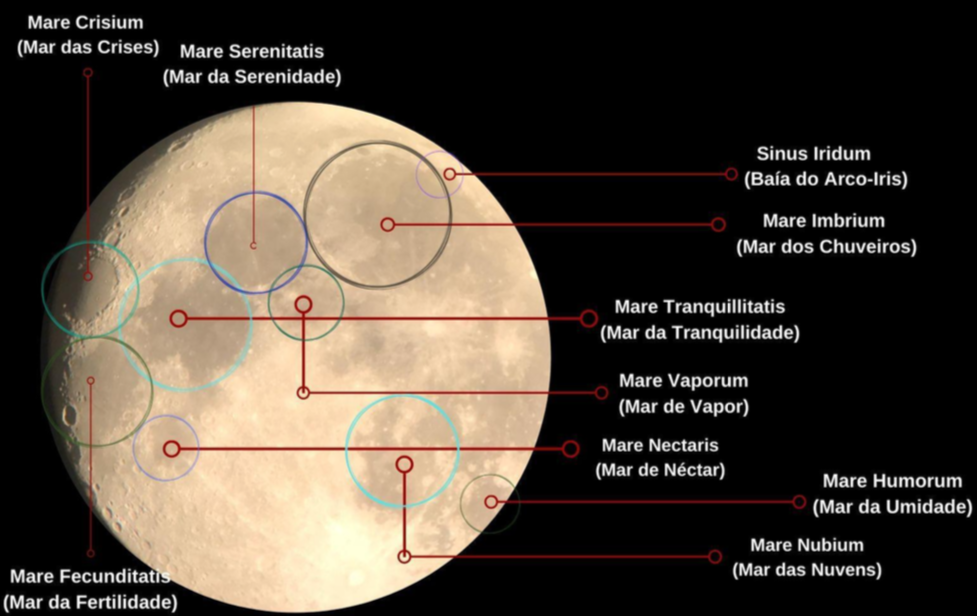}
        \end{center}
        \caption{Captura de imagem realizada no dia 13/06/2019 pelo grupo astronômico {\it{See Astronomy}}, utilizando um {\it{Smartphone}} acoplado ao telescópio Meade ETX-90 e uma lente ocular de 12mm. Nela, destacamos os mares apresentados.}
        \label{Fig:Lua1}
    \end{figure}
    
     \begin{figure}[!hbtp]
        \begin{center}
            \includegraphics[scale=0.23]{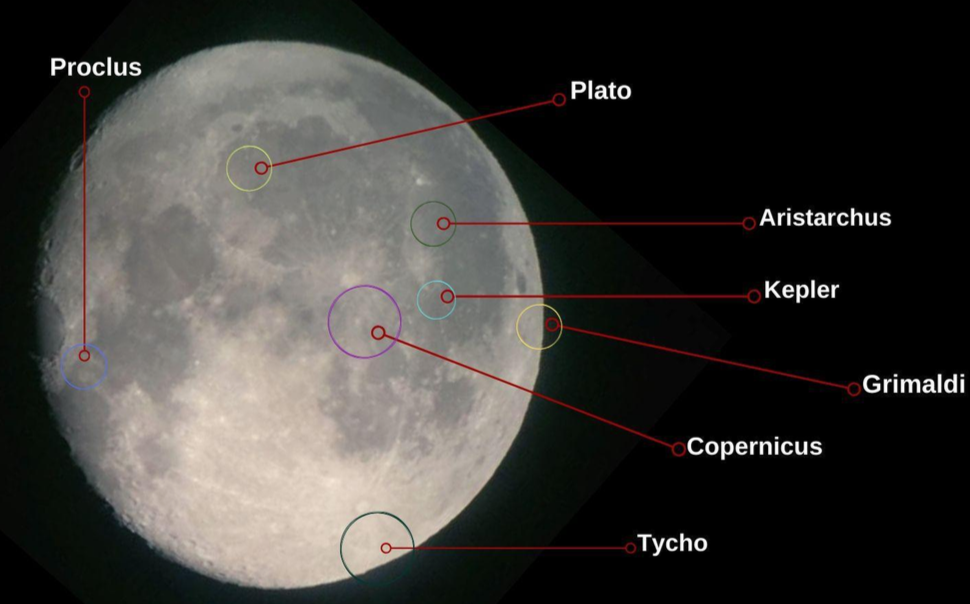}
        \end{center}
        \caption{Captura de imagem realizada no dia 14/05/2019, utilizando um {\it{Smartphone}} acoplado no telescópio Meade ETX-90 com uma lente ocular 12mm. Nela, destacamos algumas crateras na borda da superfície da Lua e algumas grandes crateras centrais.}
        \label{Fig:Lua2}
    \end{figure}
    
    \begin{figure}[!hbtp]
        \begin{center}
            \includegraphics[scale=0.335]{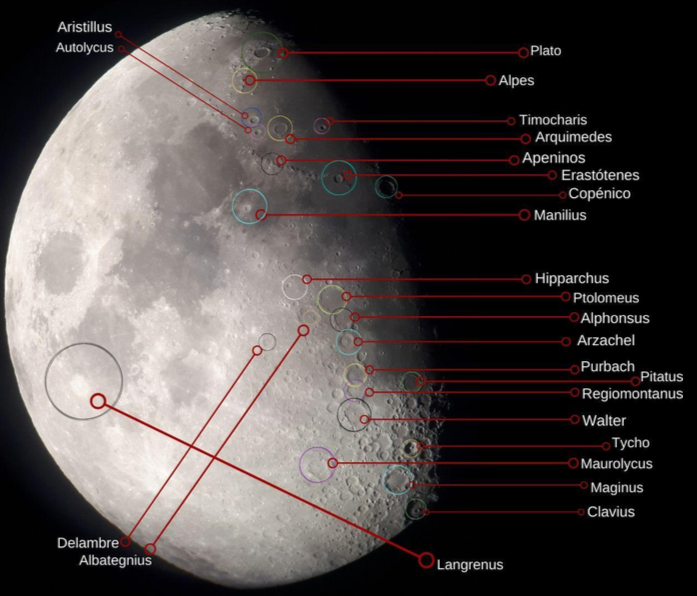}
        \end{center}
        \caption{Captura de imagem realizada no dia 09/08/2019, utilizando um {\it{Smartphone}} acoplado no telescópio Meade ETX-90 com uma lente ocular 10mm. Aqui, destacamos as crateras mais centrais da superfície lunar. Isso só foi possível devido ao contraste gerado nessa região, consequência de uma porção menos iluminada da Lua.}
        \label{Fig:Lua3}
    \end{figure}

    Em primeiro plano, para identificar as irregularidades da superfície lunar, foi abordado, primeiramente, o que seriam os  mares lunares. 
    A nomenclatura ``Mar'' refere-se às manchas escuras visíveis na Lua e foi empregada por Galileu com o seu sentido denotativo, uma vez que este acreditava em uma concreta existência de corpos d’água no astro. 
    No entanto, atualmente, o nome permanece o mesmo, ainda sabendo que os mares lunares são, ao contrário do que se acreditava, planícies basálticas oriundas de um período geologicamente ativo do satélite, onde fissuras em sua crosta expeliram lava que se solidificou, deixando-o, portanto, com grandes áreas planas com tonalidades escurecidas. 
    Dessa maneira, para identificar cientificamente estes mares, a União Astronômica Internacional (IAU, do inglês {\it{``International Astronomical Union''}}) utiliza neles nomenclaturas que fazem analogias a estados emocionais humanos - como podemos perceber no mar identificando por tranquilidade na figura (\ref{Fig:Lua1}).

    Além disso, foi importante fazermos, ainda, um estudo acerca das crateras lunares. 
    Devido à escassa gravidade que o astro apresenta, cerca de 1,62 m/s$^2$, ou seja, $1/6$ da gravidade terrestre, a colisão de corpos celestes com a superfície lunar, ao longo do tempo, foi o que levou a Lua a apresentar inúmeras deformações basálticas em sua superfície. Essas crateras, assim como os mares, possuem identificações preestabelecidas pela IAU, entretanto, seguem um critério diferente: são identificadas com nomes os quais homenageiam grandes figuras científicas e filosóficas - como é caso, respectivamente, das crateras Kepler (Johannes Kepler) e Plato (Platão), nas figuras (\ref{Fig:Lua2} e \ref{Fig:Lua3}).

    Vale ressaltar, portanto, que esse estudo poderá ser utilizado não só para a identificação das crateras e mares lunares, mas também servirá de base para relacionarmos com outros  detalhes e eventos que o astro apresenta. 
    Nesse sentido, podemos abordar as missões espaciais na Lua, por exemplo, localizando o Mar da Tranquilidade, mostrando aos estudantes que foi nessa região em que Neil Alden Armstrong deu ``um grande passo para a humanidade''. 
    Além disso, uma abordagem acerca dos movimentos de rotação da Lua (movimento ao redor de um eixo imaginário que passa pelo seu centro) e de revolução dela (movimento que realiza ao redor da Terra) poderão ser relacionados com a visualização parcial do satélite. 
    Isso foi apresentado mostrando a eles que só conseguimos observar mares e crateras lunares em uma das faces do satélite, já que a outra apresenta-se constantemente oculta devido ao sincronismo destes movimentos. 
    Essa etapa se comunica com a habilidade da BNCC

    {\it{EF04CI11} -} Associar os movimentos cíclicos da Lua e da Terra a períodos de tempo regulares e ao uso desse conhecimento para a construção de calendários em diferentes culturas.
\end{itemize}

\subsection{Construindo um Diagrama de Hertzsprung-Russell}

O poder de resolução de um instrumento óptico está relacionado com sua capacidade de separar as fontes de luz, permitindo que um observador identifique-as de forma clara. Quando isso ocorre, dizemos que este equipamento consegue resolver a imagem. Utilizando-se desse princípio, Galileu conseguiu resolver a imagem de alguns aglomerados de estrelas\footnote{Aglomerados de estrelas são estruturas estelares que interagem gravitacionalmente, para maiores detalhes sobre essas estruturas veja \url{https://pt.wikipedia.org/wiki/Aglomerado_estelar}.} em suas observações, com a luneta. 

Quando olhamos para estruturas como a Caixinha de Jóias (NGC4755), sem o uso de instrumentos ópticos, tudo que vemos é um ``pontinho luminoso''. 
A partir desta observação, seríamos levados a concluir que NGC4755 seria mais uma estrela  presente no céu.
Contudo, se fizermos o uso de um telescópio, semelhante ao de Galileu, por exemplo,  veremos  que, na verdade, NGC4755 é composto por uma infinidade de  pontos luminosos imperceptíveis ao olho humano, devido à proximidade aparente entre eles.

Uma característica interessante dessas estruturas vem do fato de elas terem sido formadas quase no mesmo momento, por isso devem apresentar semelhanças em suas composições químicas.
Com o uso da espectroscopia,  que consiste no estudo da divisão da luz com o auxílio de uma rede de difração, conseguimos evidências que comprovam tal hipótese,  visto que as linhas de absorção detectadas em seus espectros são as mesmas.
Outra informação que  pode ser obtida analisando o espectro estelar é a determinação da temperatura de uma estrela, que, por sua vez, relaciona-se com o brilho dela.
Essa relação foi descoberta independentemente por dois astrônomos,  Ejnar Hertzsprung, em 1911, e Henry Norris Russell, em 1913, após observarem que havia um padrão entre  as luminosidades e as temperaturas superficiais das estrelas, desenvolvendo, a partir desta, o diagrama H-R.
Este mostra uma relação entre a magnitude absoluta (ou luminosidade), que é um indicativo do brilho da estrela, e o tipo espectral, que está ligado à temperatura superficial desses astros.  

Assim, ao obtermos esta relação, poderemos  fazer um estudo acerca do ciclo evolutivo e classificação estelar. Dessa maneira, elaboramos uma prática que buscou obter o Diagrama H-R de um conjunto de estrelas.
A fim de atingirmos este propósito, seguimos a estratégia aplicada anteriormente, determinando os seguintes objetivos parciais:

\begin{itemize}
    \item \textbf{D1 - Reconhecendo as constelações Zodiacais}
   
   Iniciamos esse reconhecimento construindo um mapa estelar, ou planisfério celeste\footnote{Moldes do planisfério celeste, para diferentes  latitudes, podem ser encontrado no link \url{http://www.if.ufrgs.br/~fatima/planisferio/celeste/planisferio.html}},  o qual é utilizado comumente para localizar estruturas astronômicas, tais como: estrelas, planetas e chuvas de meteoros.
    Nesse sentido,  aproveitamos essa atividade para percebermos a relação entre a posição de um observador no globo terrestre e o céu observado por ele, permitindo verificar alterações nas constelações observadas quando se muda a latitude.
    Cabe entender, a princípio, que essa mudança ocorre devido à alteração da porção visível da esfera celeste, a qual está limitada pelo plano tangente à posição do observador.
    Contudo, quando mudamos a longitude de observação, apenas o horário que o astro nasce ou se põe muda, pois não ocorre mudança na altura do polo celeste, apenas o sentido de rotação da Terra que é de oeste para leste, faria com que o horário de nascimento e ocaso fossem diferentes.   
    
Por outro lado, as constelações observadas dependem também da posição da Terra com relação ao Sol. Nesse sentido, o movimento de translação muda o céu observado, já que, se o conjunto de estrelas estiver atrás do Sol, não o observaremos. 
    Isso pode ser entendido utilizando o Stellarium, mudando-se os meses e fixando o horário de observação. 

    Ainda, utilizando o simulador de céu, mostramos o caminho percorrido pelo Sol ou, de outra forma, as 13 (treze) constelações do zodíaco pelas quais nossa estrela passa ao longo do ano.
    Para tal, tiramos o efeito tanto da atmosfera quanto da superfície e, em seguida, fixamos a observação no Sol e utilizamos a função data para avançar os dias. 

    Com base no exposto acima, e tendo em mente que a posição relativa do planeta com relação ao Sol determina as regiões da Terra que recebem maior ou menor incidência dos raios solares, podemos relacionar as constelações observadas à noite com a época de calor ou frio, ou seja, cria-se uma relação entre as estações do ano e as estrelas observadas no céu. 
    A qual nos permite, inclusive, fazer previsões de quando uma das estações vai ocorrer.

    Durante a execução desta etapa, trabalhamos as seguintes habilidades previstas pela BNCC:

    {\it{ EF05CI10} - } Identificar algumas constelações no céu e os períodos do ano em que elas são visíveis;
    
    {\it{EF05CI11} - } Associar o movimento diário do Sol e das demais estrelas no céu ao movimento de rotação da Terra;
    
    {\it{EF08CI13} - } Representar os movimentos de rotação e translação da Terra e analisar o papel da inclinação do eixo de rotação da Terra em relação à sua órbita na ocorrência das estações do ano, com a utilização de modelos tridimensionais.
    
    \item \textbf{D2 - Coleta de Dados e Construindo o Diagrama H-R}
    
   Após um primeiro contato com as constelações, tido durante a atividade anterior, iniciamos a coleta de dados, que se alinha com o reconhecimento do céu e nos permite um aprofundamento das habilidades supracitadas. 
   Assim, aproveitando esta temática, verificamos que as estrelas de uma constelação não apresentam nenhum tipo de interação. 
    Elas estão aparentemente próximas devido à nossa limitação em perceber a diferença de profundidade entre as estrelas que compõe uma constelação. 
    Portanto, temos a impressão de que todas as estrelas do céu ocupam a mesma casca esférica, ou seja, estão a uma mesma distância de nós.

    Consequentemente, isso nos leva a uma questão alinhada com a obtenção do diagrama H-R: se, hipoteticamente, todas as estrelas estivessem a uma mesma distância de nós, como as veríamos?
A grandeza que expressa a hipótese acima é a magnitude absoluta, ou, de outra forma, o brilho observado por nós se todas as estrelas estivessem a uma distância de 10 pc\footnote{1pc = 206265 UA e 1 UA $\approx 1,5 \times 10^{11}$ m} (dez parsec). 

    Sabendo que a relação entre a magnitude absoluta $M$ e a luminosidade da estrela $L$ é dada por:
    $$ M= C' - 2,5\log{L} + 5 \quad,$$
    sendo $C'$ uma constate que depende dos ajustes de escala, concluímos que quanto maior a magnitude absoluta da estrela, menor deve ser a luminosidade dela. Para maiores detalhes sobre como foi obtida a relação acima, veja a referência \cite{comins2010descobrindo}.

    Como estávamos interessados em obter a relação entre a luminosidade (ou magnitude absoluta, pela equação acima) e a temperatura (obtida via tipo espectral ou índice de cor B-V), iniciamos, então, a coleta desses dados com auxilio do Stellarium. 
    Ademais, coletamos informações como a distância das estrelas e suas identificações de acordo com o Catálogo Hipparcos\footnote{O Catálogo Hipparcos pode ser encontrado em \url{https://www.cosmos.esa.int/web/hipparcos}}, com a finalidade de ampliar o conhecimento e reafirmar a abordagem acima.
    Sendo assim:

    (i)  - Escolhemos um conjunto de constelações, dando preferência às do zodíaco e outras mais conhecidas, como Órion, Cruzeiro do Sul, Centauro e Cão Maior;
    
    (ii) - Elaboramos uma listagem de controle, de modo que o maior número possível de constelações tivessem suas informações coletadas;

    (iii) - Criamos uma tabela compartilhada, de modo que todos os colaboradores pudessem lançar os dados simultanemante;
    
    (iv) - Identificamos as constelações no Stellarium e anotamos os dados das estrelas mais brilhantes, as que formam sua figura.
    
    (v) - Plotamos os pontos em um plano cartesiano,  tal que o  Tipo Espectral e a  Magnitude Absoluta correpondam aos eixos X e Y, respectivamente.
Para isso, utilizamos a seguinte correspondência numérica para os tipos espectrais: O=10, B=20, A=30, F=40,  G=50, K=60 e  M=70. 
Desse modo, um conjunto de estrelas com os  tipos espectrais A2, B6, M0 e K5.5, por exemplo,  assumem os valores de 32, 26, 70 e 65.5, respectivamente,  dentro dessa definição.

Com base nisso, o resultado mais provável obtido usando os dados coletados será algo como mostra a figura (\ref{Fig:DiagHR}).
Ou seja, nenhuma relação entre a magnitude absoluta e a classe espectral será observada.
    \begin{figure}[!hbtp]
        \begin{center}
            \includegraphics[scale=0.4]{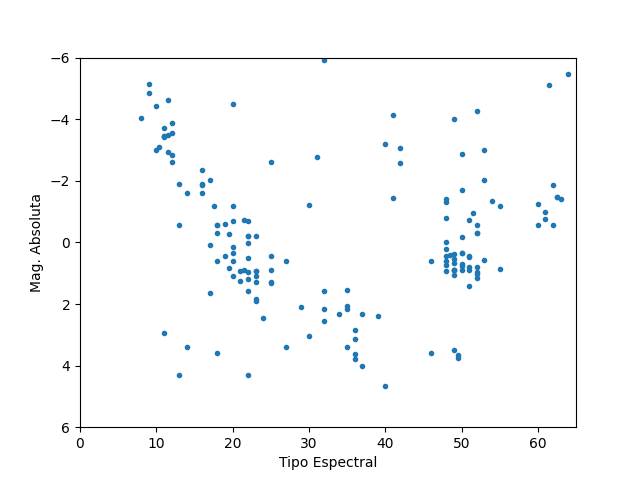}
        \end{center}
        \caption{Diagrama H-R construído com todos os dados. Observe que há uma grande dispersão dos dados.}
        \label{Fig:DiagHR}
    \end{figure}
     Por outro lado, quando limitamos a coleta de dados para estrelas próximas, distâncias menores que $5$ pc, encontramos a curva mostrada na figura (\ref{Fig:DiagHR5pc}).

    \begin{figure}[!hbtp]
        \begin{center}
            \includegraphics[scale=0.4]{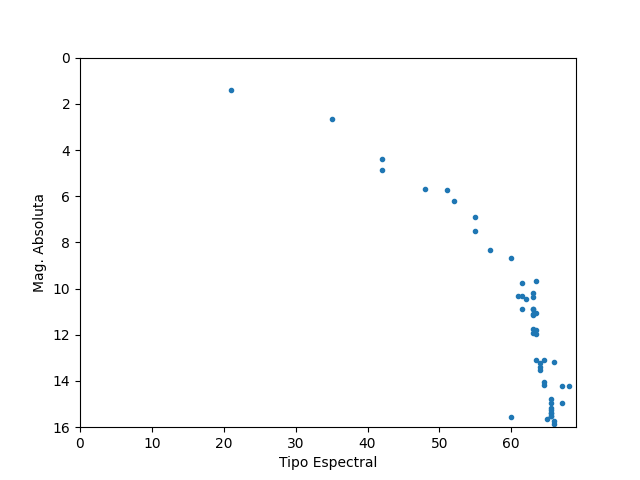}
        \end{center}
        \caption{Diagrama H-R construído com todos os dados de estrelas cujas distâncias são inferiores a $5$pc. Observe que as estrelas estão na sequência principal.}
        \label{Fig:DiagHR5pc}
    \end{figure}

    Neste caso, as estrelas apresentam uma tendência a ocupar uma região muito particular nesse diagrama, chamada de sequência principal.
    Vale destacar que este diagrama permite determinar o estágio evolutivo que se encontra a estrela.
    No caso particular apresentado na figura, onde as estrelas estão na sequência principal, é permitido concluir que a estrela mantém sua estrutura devido ao equilíbrio hidrostático entre a força gerada pela pressão do gás/plasma\footnote{Para maiores detalhes sobre a relevância da pressão do plasma no equilíbrio hidrostático acesse \url{http://astro.if.ufrgs.br/evol/node13.htm}.} e a força gravitacional.
    Outra evidência que pode ser comprovada quando aumentamos a quantidade de dados coletados é que a maioria das estrelas vistas à noite se encontram neste local do gráfico.
\end{itemize}

\section{CONCLUS\~AO}

A deficitária formação dos profissionais que trabalham com a temática de Astronomia no  ensino básico pode trazer prejuízos à formação intelectual dos estudantes, nos tópicos previstos pela BNCC.
Sendo assim, torna-se necessário ampliar o apoio técnico para que o conhecimento seja transmitido de forma mais adequada. 
Uma  alternativa a essa problemática é a formação continuada que tem sido implementada, como \cite{oliveira2020, bartelmebs2016ensino, langhi2018formaccao}. 
Nesse sentido, buscamos construir três sequências didáticas que podem ser aplicadas tanto nos espaços formais quanto nos não formais de ensino utilizando o Stellarium, as quais se alinham com as habilidades da BNCC e com os planos de trabalho dos bolsistas de IC-Jr do OAIG. 
Embora essa SD esteja majoritariamente voltada à educação básica e, por isso, tende a se comunicar com os currículos desse nível de ensino, ela pode ser ajustada para atender, até mesmo, alunos de graduação.

Isso permitiu que cada bolsista pudesse atingir os objetivos previstos em suas pesquisas e, ainda, elaborar três práticas didáticas: determinação da massa de Júpiter, identificação dos mares e crateras lunares, além da construção de um diagrama H-R.
Por conseguinte, a aplicação dessas práticas permitiu que trabalhássemos algumas habilidades da BNCC, tais como: o reconhecimento de algumas estruturas celestes, as condições para ocorrência de eclipses, os efeitos dos astros sobre a Terra, a relação entre a luminosidade e o tipo espectral, bem como a aplicabilidade da gravitação newtoniana para obtenção da massa de Júpiter.
Paralelamente, a construção do diagrama H-R mostrou ao estudante que a seleção dos dados coletados é um fator importante na experimentação.
Sendo assim, as três temáticas aproximam o aluno do conhecimento científico e  propõem uma interação com a natureza, levando-os à resultados satisfatórios, uma vez que a massa de Júpiter obtida foi de $ M_J = (1,9 \pm 0,5)\times 10^{27}$ kg, as fotos da Lua permitiram o reconhecimento de um número razoável de crateras e mares lunares, e conseguimos obter o diagrama H-R para um conjunto de estrelas que estão a menos de $5$ pc de distância, como mostrado na figura (\ref{Fig:DiagHR5pc}).
Contudo, ainda podemos pensar em outras problemáticas, alinhadas a esta SD, que podem ser trabalhadas futuramente, tais como: (i) a determinação da velocidade da luz, a partir da diferença no tempo medido para a duração do eclipse de Io quando a Terra está mais próxima de Júpiter e quando ela está mais afastada do planeta e (ii) a obtenção do diagrama H-R para estrelas mais distantes.

O uso destas sequências didáticas se mostrou um grande aliado na tratativa da temática proposta, trazendo como ponto positivo o respeito ao tempo de aprendizagem de cada aluno, permitindo-lhe o empoderamento do conhecimento, tornando-o um agente ativo e protagonista do processo ensino-aprendizagem.
Um dos fatores que comprovam tal conjuntura é a perceptível evolução na segurança em discutir as temáticas aqui apresentadas, bem como na capacidade de conduzir as intervenções com os visitantes do OAIG, possibilitando o exercício do protagonismo estudantil.

\section{Agradecimentos}
Os autores agradecem \`a Fapes pelo financiamento do projeto, ao Ifes e CNPq pelo apoio com bolsas, ao Cosmo-Ufes e Ufmg pelo apoio técnico e aos alunos de ensino médio ligados ao Observatório Astronômico do Ifes Guarapari, que ajudaram com a coleta de dados e discussões.

\begin{sobreautor}[m]
Adriano Mesquita Oliveira (adriano.oliveira@ifes.edu.br) é Doutor em Física pela {\it{Universidade Federal do Espírito Santo}} (Ufes) e: (1) atua como professor de Física do {\it{Instituto Federal do Espírito Santo}} (Ifes-Guarapari), onde ministra aulas para os ensinos médio e superior; (2) coordena o Observatório Astronômico do Ifes Guarapari (OAIG) e o curso de Formação Continuada para Professores do Ensino Fundamental; (3) está como diretor da Diretoria de Pesquisa, Pós-Graduação e Extensão o Ifes Guarapari e (4) faz parte da equipe executora da Mostra de Astronomia do ES, Encontro de Astronomia do ES.

Cibele Kemeicik (cikemeicik@gmail.com) é mestranda do Programa de Pós-Graduação em Ensino de Física do Instituto Federal do Espírito Santo (PPEFis) e atua como professora da rede municipal de Guarapari e Anchieta, onde ministra aulas de ciências.

Augusto C. T. Monteiro (augusto.monteiro@ifes.edu.br) é Mestre em Matemática pela {\it{Universidade Federal do Espírito Santo}} (Ufes) e Professor de Matemática do {\it{Instituto Federal do Espírito Santo}} (Ifes-Guarapari), onde  ministra aulas para os ensinos médio e superior. Além disso, é o vice-coordenador do OAIG e do curso de Formação Continuada para Professores do Ensino Fundamental e está como Coordenador de Extensão do Ifes Guarapari.

Thalita S. Benincá (thalitasartoribeninca@gmail.com) é aluna do terceiro ano do Curso Técnico, Integrado ao Ensino Médio, em  Administração e atua como monitor do OAIG.

Carlos Daniel da S. Mattos (carlosdanieldasilva703@gmail.com) é aluno do terceiro ano do Curso Técnico, Integrado ao Ensino Médio, em Administração  e atua como monitor do OAIG.

Guilherme L. Schmidt (guilherme220403@hotmail.com) é aluno do terceiro ano do Curso Técnico, Integrado ao Ensino Médio, em Mecânica e atua como monitor do OAIG.
\end{sobreautor}

\end{document}